\documentclass{ws-procs9x6}

\begin{document}

\title{Unified semiclassical approach to isoscalar collective modes
in heavy nuclei}

\author{V. I. ABROSIMOV}

\address{Institute for Nuclear Research, 03028 Kiev, Ukraine \\
E-mail: abrosim@kinr.kiev.ua}

\author{A. DELLAFIORE and F. MATERA}

\address{Istituto Nazionale di Fisica Nucleare  and Dipartimento di Fisica,\\
 Universit\`a di Firenze, via Sansone 1, 50019 Sesto F.no (Firenze), Italy
 \\
E-mail:della@fi.infn.it; matera@fi.infn.it}

\maketitle

\abstracts{
A semiclassical model based on the solution of the Vlasov
equation for finite systems with a sharp moving surface has been used
to study the isoscalar quadrupole and octupole collective modes in heavy
spherical nuclei. Within this model, a unified description
of both low-energy surface modes and higher-energy giant resonances
has been achieved by introducing a coupling between surface
vibrations and the motion of single nucleons.
Analytical expressions for the collective response functions of
different multipolarity can be derived by using a separable
approximation for the residual interaction between nucleons.
The response functions obtained in this way give a good qualitative
description of the quadrupole and octupole response in heavy nuclei.
Although shell effects are not explicitly included in the theory, our
semiclassical response functions are  very similar to the
quantum ones. This happens because of the well known close relation
between classical trajectories and shell structure.
The role played by particular nucleon trajectories and their
connection with various features of the nuclear response is displayed
most clearly in the
present approach, we discuss in some detail the damping of low-energy
octupole vibrations and give an explicit expression showing that only
nucleons moving on triangular orbits can contribute to this damping.}

\section{Introduction}
It is well known that the isoscalar quadrupole and octupole response of
nuclei displays both low- and high-energy collective modes \cite{vdw}.
Also known is that semiclassical models have difficulties in describing both
these systematic features of the isoscalar response, in particular, models based
on fluid dynamics, see e.g. \cite{holz}, can explain the giant resonances, but
fail to describe the low-energy collective modes. On the other hand it is known
from quantum studies that the coupling between the motion of individual nucleons
and surface vibrations plays an essential role in low-energy nuclear collective
modes, see e.g. \cite{bort,b&b,lac}. Semiclassical models of the fluid-dynamical
type do not contain explicitly the single-particle degrees of freedom, so they
can not describe the coupling between individual nucleons and surface motion.

In the present contribution
we review a study the isoscalar collective modes in nuclei \cite{adm4,adm5}
made by using a semiclassical approach that includes the
single-particle degrees of freedom explicitly and thus
allows for an account the coupling between
individual nucleons and surface motion.
Our model is based on the linearized
Vlasov kinetic equation for finite systems with moving surface \cite{bri,ads}.
The coupling between the motion of individual nucleons and the
surface vibrations is obtained by treating the nuclear surface as a
collective dynamical variable, like in the liquid drop model.
Here we concentrate  our attention on the
isoscalar quadrupole and octupole collective modes in heavy spherical
nuclei, an application of the same model to the compression dipole
modes has been discussed in a previous meeting of this series \cite{maiori}.

\section{Reminder of formalism}
This Section recalls briefly the formalism of References
\cite{bri,ads} which is at the basis of the present approach.

The fluctuations of the phase-space density induced by a weak
external force can be described by the linearized Vlasov equation,
which is usually a differential equation in seven variables. For spherical
systems this equation can be reduced to a system of two (coupled)
differential equations in the radial coordinate alone \cite{bri}.
This is achieved by means of a change of variables and a partial-wave
expansion:
\begin{eqnarray}
&&\delta f({\bf r}, {\bf p},\omega)=\sum_{LMN}[\delta
f^{L+}_{MN}(\epsilon,\lambda,r,\omega) +
\delta f^{L-}_{MN}(\epsilon,\lambda,r,\omega)]\nonumber\\
&&\times {\Big (} {\mathcal D}^{L}_{MN}
(\alpha,\beta,\gamma){\Big)}^{*} Y_{LN}(\frac{\pi}{2}\frac{\pi}{2})\,.
\end{eqnarray}
The functions $\delta f^{L\pm}_{MN}(\epsilon,\lambda,r,\omega)$ are
partial-wave components of the (Fourier
transformed in time) density fluctuations for particles with energy
$\epsilon$, magnitude of angular momentum $\lambda$ and radial position $r$,
the $\pm$ sign  distinguishes between particles having
positive or negative components of the radial momentum $p_{r}$.

The other terms in the expansion are Wigner matrices and spherical
harmonics.

In order to solve the one-dimensional linearized Vlasov equation
for the $\delta f^{L\pm}_{MN}$ functions we must specify
the boundary conditions satisfied by these functions.
Different boundary conditions allow us to study different physical
properties of the system, so the fixed-surface boundary conditions
employed in \cite{bri} were adequate to study giant resonances, but
different (moving-surface) boundary conditions \cite{ads} must be introduced
in order to study surface modes.  We assume that the external force can also
induce oscillations of the system surface according to the usual liquid-drop
model expression
\begin{equation}
\label{rot}
R(\vartheta,\varphi,t)=R+\sum_{LM} \delta R_{LM}(t)
Y_{LM}(\vartheta,\varphi)
\end{equation}
and the boundary condition satisfied by the functions
$\delta f^{L\pm}_{MN}$ at the nuclear surface is taken as
\begin{equation}
\label{msbc}
\delta f^{L+}_{MN}(R)-\delta f^{L-}_{MN}(R)=2F'(\epsilon)i\omega
p_{r}\delta R_{LM}(\omega)\,.
\end{equation}
This equation has been derived with the assumption that the equilibrium
phase-space density is a function $F(\epsilon)$ of the particle energy
alone, $F'(\epsilon)$ is its derivative.
The boundary condition (\ref{msbc}) corresponds to a mirror reflection
of particles in the reference frame of the moving nuclear surface, 
it provides a coupling
between the motion of nucleons and the surface vibrations.
A self-consistency condition involving the nuclear surface tension is
then used to determine the time (or frequency) dependence of the additional
collective variables $\delta R_{LM}$(t) \cite{ads}.

Now, assuming a simplified residual interaction of separable form,
\begin{equation}
\label{vres}
v(r_{1},r_{2})=\kappa_{L} r_{1}^{L}r_{2}^{L}\,,
\end{equation}
the moving-surface isoscalar collective response function of a
spherical nucleus,
described as a system of $A$ interacting nucleons contained in
a cavity of equilibrium radius $R=1.2A^{\frac{1}{3}}\: {\rm fm}$, is given
by
\begin{equation}
\label{rtil}
\tilde{{\mathcal R}}_{L}(s)=
{\mathcal R}_{L}(s)+ {\mathcal S}_{L}(s)\,.
\end{equation}
Instead of the frequency $\omega$, as independent variable we have
used the more convenient dimensionless quantity $s=\omega/(v_{F}/R)$
($v_{F}$ is the Fermi velocity). The  response
function ${\mathcal R}_{L}(s)$, given by
\begin{equation}
\label{collfix}
{\mathcal R}_{L}(s)=\frac{{\mathcal R}_{L}^{0}(s)}{1-\kappa_{L} {\mathcal
R}_{L}^{0}(s)}\,,
\end{equation}
describes the collective response in the fixed-surface limit.
The response function ${\mathcal R}_{L}^{0}(s)$ is analogous
to the quantum single-particle response function  and
it is given explicitly by \cite{adm4,adm5},
\begin{equation}
\label{pi0}
{\mathcal R}_{L}^{0}(s)=
\frac{9A}{8\pi}\frac{1}{\epsilon_{F}}\sum_{n=-\infty}^{+\infty}
\sum_{N=-L}^{N=L}(C_{LN})^{2} \int_{0}^{1} dx x^{2}
~s_{nN}(x){{(Q^{(L)}_{nN}(x))^{2}}\over {s+i\varepsilon -s_{nN}(x)}}\,,
\end{equation}
where $\epsilon_{F}$ is the Fermi energy and the quantity $\varepsilon$ is
a vanishingly small parameter that determines the integration path at poles.

The functions $s_{nN}(x)$ are defined as
\begin{equation}
s_{nN}(x)=\frac{n\pi +N\arcsin(x)}{x}.
\end{equation}
The variable $x$ is related to the classical nucleon angular momentum
$\lambda$.
The quantities $C_{LN}$ in Eq. (\ref{pi0}) are classical limits of the
Clebsh-Gordan coefficients coming from the angular integration.
In principle the integer $N$ takes values between $-L$ and $L$, however
only the coefficients $C_{LN}$ where $N$ has the same parity as $L$
are nonvanishing.
The coefficients $Q^{(L)}_{nN}(x)$ appearing in the numerator of Eq.
(\ref{pi0}) have been defined in Ref. \cite{bri}, they are essentially the
classical limit of the radial matrix elements of the multipole operator
$r^{L}$ and can be evaluted analytically for $L=2,3$.

The function ${\mathcal S}_{L}(s)$ in Eq. (\ref{rtil}) gives the
moving-surface contribution to the response. With the simple
interaction (\ref{vres})
this function can be evaluated explicitly as \cite{adm4,adm5}
\begin{equation}
\label{surfresp}
{\mathcal S}_{L}(s)=-\frac{R^{6}}{1-\kappa_{L} {\mathcal
R}_{L}^{0}(s)}\: \frac{[\chi^{0}_{L}(s)+\kappa_{L}\varrho_{0}R^{L}
{\mathcal R}_{L}^{0}(s)]^{2}}{[C_{L}-\chi_{L}(s)][1-\kappa_{L} {\mathcal
R}_{L}^{0}(s)]+\kappa_{L}R^{6}[\chi^{0}_{L}(s)+\varrho_{0}R^{L}]^{2}}\,,
\end{equation}
with
$C_{L}=\sigma R^{2}(L-1)(L+2)+(C_{L})_{coul}$,
$\sigma\approx 1 {\rm MeV\,fm^{-2}}$ is the surface tension
parameter obtained from the mass formula, $(C_{L})_{coul}$ gives
the Coulomb contribution to the restoring force and
$\varrho_{0}=A/\frac{4\pi}{3}R^{3}$ is the equilibrium density.
The functions $\chi^{0}_{L}(s)$ and $\chi_{L}(s)$
are given by \cite{adm2}
\begin{equation}
\label{chik}
\chi^{0}_{L}(s) =\frac{9A}{4\pi}\frac{1}{R^{3}}\sum_{n=-\infty}^{+\infty}
\sum_{N=-L}^{N=L}(C_{LN})^{2}\int_{0}^{1} dx x^{2}~
s_{nN}(x){(-)^{n}{Q^{(L)}_{nN}(x)}
\over {s+i\varepsilon-s_{nN}(x)}},
\end{equation}
and
\begin{equation}
\label{chick}
\chi_{L}(s)=-\frac{9A}{2\pi}\epsilon_{F}\,(s+i\varepsilon)
\sum_{n N} (C_{LN})^{2} \int_{0}^{1} dx x^{2}~ \frac{1}
{s+i\varepsilon-s_{nN}(x)}\,,
\end{equation}
their structure is similar to that of the zero-order propagator (\ref{pi0}).

We refer to the papers \cite{adm4,adm5} for further details on the formalism
and discuss here only the main points.

Equation (\ref{surfresp}) is the main result in the present context.
Together with Eqs.(\ref{rtil}) and (\ref{collfix}), this equation
gives a unified expression of the isoscalar response function, including
both the low- and high-energy collective excitations.
By comparing the fixed- and moving-surface response functions, we can
appreciate the effects due to the coupling between the motion
of individual nucleons and the surface vibrations.

\section{Fixed- vs. moving-surface strength distributions}
The strength function $S_{L}(E)$ associated with the
response function (\ref{rtil}) is defined as
($E=\hbar \omega$)
\begin{equation}
\label{sfl}
S_{L}(E)=-\frac{1}{\pi}{\rm Im}\,\tilde{{\mathcal R}}_{L}(E)\,.
\end{equation}
We discuss here the isoscalar quadrupole and octupole strength distributions.
The strength $\kappa_{L}$ of the residual interaction (\ref{vres}) can be
estimated in a self-consistent way, giving (\cite{bm2}, p. 557),
\begin{equation}
\kappa_{BM} =-\frac{4\pi}{3}\frac{m\omega_{0}^{2}}{A R^{2L-2}},
\end{equation}
with the parameter $\omega_{0}$ given by
$\omega_{0}\approx 41 A^{-\frac{1}{3}}{\rm MeV}$.
Since this estimate is based on a harmonic oscillator mean field and
we are assuming a square-well potential instead, we expect some
differences. Hence we  determine the parameter $\kappa_{L}$
phenomenologically, by requiring that the peak of the high-energy
resonance agrees with the experimental value of the giant multipole
resonance energy. This requirement implies
$\kappa_{L} \approx 2\kappa_{BM}$ for $L=2,3$.

In Fig.1 we display the quadrupole strength function (L=2 in
Eq. (\ref{sfl})) obtained for $A=208$ nucleons by using different
approximations.
The dotted curve is obtained from the zero-order response function
(\ref{pi0}), it is similar to the quantum response evaluated in the
Hartree-Fock approximation. The dashed curve is obtained from the
collective fixed-surface response function (\ref{collfix}). Comparison
with the dotted curve clearly shows the effects of collectivity.
The collective fixed-surface response has one giant quadrupole peak.
Our result for this peak is very similar to that of the
recent random-pase approximation (RPA) calculations of \cite{ham}
(cf. Fig.5 of \cite{ham}). However, contrary to the RPA calculations,
there is no signal of a low-energy peak in the fixed-surface response function.
The solid curve instead shows the
moving-surface response given by Eqs. (\ref{rtil}) and (\ref{surfresp}).
Now a broad bump appears in the
low-energy part of the response and a narrower peak is still present
at the giant
resonance energy. Of course the details of the low-energy excitations
are determined by quantum effects, nonetheless the present semiclassical
approach does reproduce the average behaviour of this systematic
feature of the quadrupole response.

We finally notice that the width of the giant quadrupole resonance is
underestimated by our approach, this is a well known limit of
all mean-field calculations that include only Landau damping. A more
realistic estimate of the giant-resonance width would require
including a collision term into our kinetic equation.

\begin{figure}[ht]
\centerline{\epsfxsize=3.1in\epsfbox{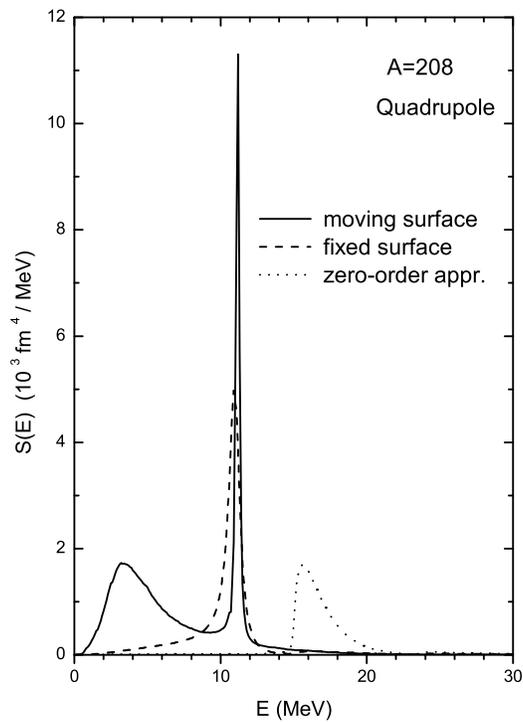}}
\caption{Quadrupole strength function for a hypothetical nucleus
of $A=208$ nucleons. The dotted curve shows the zero-order aproximation,
the dashed curve instead shows the collective response evaluated in the
fixed-surface approximation.
The full curve gives the moving-surface response.}
\end{figure}


In Fig.2 we show the octupole strength function ($L=3$ in Eq.
(\ref{sfl})). The zero-order octupole strength function (dotted curve)
is concentrated in two regions around 8 and 24 MeV.
In this respect our semiclassical response is strikingly similar to the
quantum response, which is concentrated in the $1\hbar\omega$ and
$3\hbar\omega$ regions. This concentration of strength is quite remarkable
because our equilibrium phase-space density, which is taken to be of the
Thomas-Fermi type, does not include any shell effect, however
we still obtain a strength distribution that is very similar to the one
usually interpreted in terms of transitions between different shells.

We can clearly see that the collective fixed-surface response given by
Eq. (\ref{collfix}) (dashed curve) has two sharp peaks around 20 Mev and
6-7 Mev. The experimentally observed
\cite{vdw} concentration of isoscalar octupole strength in the two regions
usually denoted by HEOR (high energy octupole resonance) and LEOR (low
energy octupole resonance) is qualitatively reproduced, however the
considerable strength experimentally observed at lower energy (low-lying
collective states) is absent from the fixed-surface response function.
The most relevant change induced by the moving surface (solid curve
in Fig.2) is the large double hump appearing at low energy. This feature is
in qualitative agreement both with experiment \cite{vdw} and with the result
of RPA-type calculations (see e.g. \cite{ll}). We interpret
this low-energy double hump as a superposition of surface vibration
and LEOR.

The moving-surface octupole response of Fig. 2 displays also a novel
resonance-like structure between the LEOR and the HEOR (at about 13
MeV for a system
of $A=208$ nucleons).

\begin{figure}[ht]
\centerline{\epsfxsize=3.1in\epsfbox{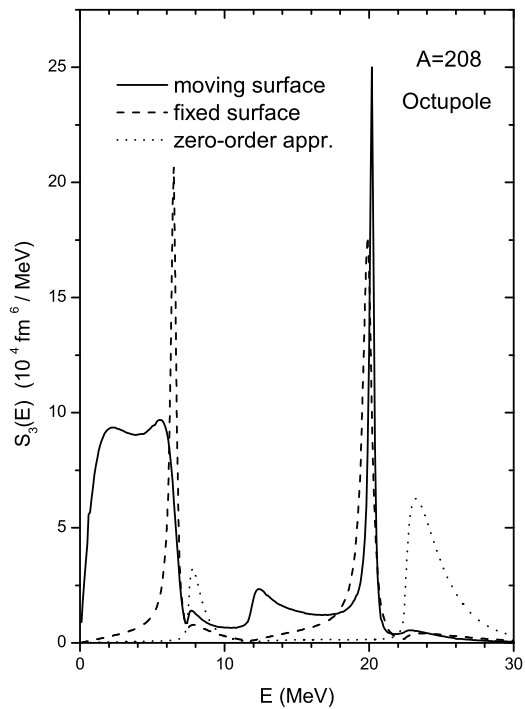}}
\caption{The same as in Fig.1 for the octupole strength function.}
\end{figure}


We also find \cite{adm1} that the parameters $\delta R_{3M}(t)$,
describing the octupole surface vibrations in Eq. (\ref{rot}),
approximately satisfy an equation of motion of the damped oscillator kind:
\begin{equation}
\label{odos}
D_{3}\delta \ddot{R}_{3M}(t) +\gamma_{3} \delta
\dot{R}_{3M}(t)+C_{3}\delta R_{3M}(t)=0\,.
\end{equation}
The friction coefficient $\gamma_{L}$ can be evaluated
analyticallty in the low-frequency limit, giving (for a generic $L$)
\begin{equation}
\label{gamma}
\gamma_L =\gamma_{wf}~2{(4\pi)^2\over{2L+1}}\sum_{N=1}^L {1\over N}
\mid Y_{LN}({\pi\over 2},{\pi\over 2})\mid^2
\sum_{n=1}^{+\infty} \cos\alpha_{nN}\sin^3 \alpha_{nN}
\Theta({\pi\over 2}-\alpha_{nN})\,,
\end{equation}
with $\gamma_{wf}=\frac{3}{4}\varrho_{0}p_{F}R^{4}$ and
$\alpha_{nN}=\frac{n}{N}\pi$. The angles $\alpha_{nN}$ are related to the
nucleon trajectories. In the octupole case the coefficient $\gamma_{L=3}$ gets
a contribution only from the term with $n=1$ and $N=3$, thus we see that
only nucleons moving along closed triangular trajectories can contribute to
the damping of surface octupole vibrations.

\section{Conclusions}
A unified description of the low- and high-energy isoscalar
collective quadrupole
and octupole response has been achieved by using appropriate
boundary conditions for the fluctuations of the phase-space density
described by the linearized Vlasov equation.
The response functions obtained in this way give a good qualitative
description of all the main features of the isoscalar
response in heavy nuclei, i. e. low-lying quadrupole and octupole
collective modes, plus quadrupole and octupole giant resonances.
In our model the low-energy modes are surface
oscillations and the coupling between single-particle motion and surface
vibrations is described by simple analytical expressions.


\begin{thebibliography}{99}
\bibitem{vdw}A. van der Woude, {\it Progr. Part. Nucl. Phys.} {\bf 18},
217 (1987).

\bibitem{holz} G. Holzwarth and G. Eckart, {\it Nucl. Phys.} {\bf A325},
1 (1979).

\bibitem{bort}P. F. Bortignon and R. A. Broglia, {\it Nucl. Phys.} {\bf A371},
405 (1981).

\bibitem{b&b}G. F. Bertsch and R. A. Broglia, {\it Oscillations in finite
quantum systems, ch.6} (Cambridge University Press, Cambridge, UK, 1994).

\bibitem{lac}D. Lacroix, S. Ayik and Ph. Chomaz, {\it Phys. Rev.} {\bf C63},
064305 (2001).

\bibitem{adm4}V. I. Abrosimov, A. Dellafiore and F. Matera,
{\it Nucl. Phys.} {\bf A717}, 44 (2003).

\bibitem{adm5}V. I. Abrosimov, O. I. Davidovskaya, A. Dellafiore and
F. Matera, {\it Nucl. Phys.} {\bf A727}, 220 (2003).

\bibitem{bri}D. M. Brink, A. Dellafiore and M. Di Toro, {\it Nucl.Phys.}
{\bf A456}, 205 (1986).

\bibitem{ads}V. Abrosimov, M. Di Toro and V. Strutinsky,
{\it Nucl.Phys.} {\bf A562}, 41 (1993).

\bibitem{maiori}V. I. Abrosimov, A. Dellafiore and F. Matera,
in {\it Proc. of the 7th Intern. Spring Seminar on Nucl. Physics},
edited by A.Covello (World Scientific, Singapore, 2002), p.481.

\bibitem{adm2} V. I. Abrosimov, A. Dellafiore and F. Matera,
{\it Nucl. Phys.} {\bf A697}, 748 (2002).

\bibitem{bm2}A. Bohr and B. M. Mottelson, {\it Nuclear Structure,
Vol. 2} (W.A. Benjamin, Inc.: Reading, Massachussets, 1975).

\bibitem{ham}I. Hamamoto, H. Sagawa and X. Z. Zhang,
{\it Nucl. Phys.} {\bf A648}, 203 (1999).

\bibitem{ll}K. F. Liu, H. Luo, Z. Ma, Q. Shen and S. A. Moszkowski,
{\it Nucl. Phys.} {\bf A534}, 1 (1991).

\bibitem{adm1}V. Abrosimov, A. Dellafiore and F. Matera,
{\it Nucl. Phys.} {\bf A653}, 115 (1999).

\end{thebibliography}
\end{document}